\documentclass[12pt,letterpaper]{JHEP}

\usepackage{epsfig}

\def\gsim{\;\raise0.3ex\hbox{$>$\kern-0.75em\raise-1.1ex\hbox{$\sim$}}\;}
\def\lsim{\;\raise0.3ex\hbox{$<$\kern-0.75em\raise-1.1ex\hbox{$\sim$}}\;}

\def\pref#1{(\ref{#1})}
\def\ignore#1{}  
\def\roughly#1{\mathrel{\raise.3ex\hbox{$#1$%
\kern-.75em\lower1ex\hbox{$\sim$}}}}
\def\lsim{\roughly<}
\def\gsim{\roughly>}

\def\eq{\begin{equation}}
\def\eeq{\end{equation}}
\def\eqa{\begin{eqnarray}}
\def\eeqa{\end{eqnarray}}
\def\ms{M_s}
\def\mpl{M_p}

\def\dperp{{d_\perp}}

\def\eps{\epsilon}
\title{Brane-Antibrane Inflation in Orbifold and Orientifold Models}

\author{C.P. Burgess,$^{1}$ P. Martineau,${}^{1}$
F. Quevedo,$^2$ G. Rajesh,$^3$ R.-J. Zhang$^4$\\

          $^1$ Physics Department, McGill University,
               3600 University Street,\\
               Montr\'eal, Qu\'ebec,  H3A 2T8 CANADA.\\
          $^2$ Centre for Mathematical Sciences, DAMTP,
               University of Cambridge,\\
               Cambridge CB3 0WA UK.\\
          $^3$ Enrico Fermi Institute and Department of Physics,
               University of Chicago,\\
               5640 Ellis Ave., Chicago IL 60637 USA.\\
          $^4$ School of Natural Sciences,
               Institute for Advanced Study,\\
               Princeton NJ 08540 USA.\\
          }

\abstract{
We analyse the cosmological implications of brane-antibrane systems
in string-theoretic orbifold and orientifold models. In a class of realistic
models, consistency conditions require branes and antibranes to be
stuck at different fixed points, and so their mutual attraction
generates a potential for one of the radii of the underlying torus
or the 4D string dilaton. Assuming that all other moduli have been
fixed by string effects, we find
that this potential leads naturally to a period of cosmic inflation with
the radion or dilaton
 field as the inflaton. The slow-roll conditions are satisfied
 more generically than if the branes were free to move within the
space. The appearance of tachyon fields at certain points in moduli space
indicates the onset of phase transitions to different non-BPS brane systems,
providing ways of ending inflation and reheating the corresponding observable
brane universe. In each case we find  relations between the
inflationary parameters and the string scale to get the correct
spectrum of density perturbations. In some examples the small 
numbers required as inputs are no smaller than 0.01, and are
the same small quantities which are required to explain the gauge
hierarchy.
}

\preprint{DAMTP-2001-97, EFI-01-47, McGill-01/23, hep-th/0111025}

\begin{document}

\section{Introduction}
Recent observations are beginning to redundantly test the standard
Big-Bang picture of late-time cosmology, and the success of this
theoretical description is determining the values which are
required for the various cosmological parameters. The success
sharpens the need to understand the puzzling initial conditions of
this picture, which are typically cast as various `problems'---such
as the flatness, horizon and cosmological-constant problems.

Inflationary cosmology \cite{Guth} represents the potentially
cleanest explanation of many of these initial-condition
problems---particularly in its hybrid
variant \cite{hybridinflation}---since it recasts the special initial
conditions of late-time
cosmology as the robust consequences of earlier dynamics which
results in the dramatic expansion of the universe. This
explanation predicts properties for the temperature fluctuations
in the cosmic microwave background (CMB), and quite generically
successfully describes the spectrum found by extant observations.
There are problems, however, and inflation remains a successful
cosmology seeking a convincing theoretical realization. Although
considerable latitude exists in model building, since the physics
of the relevant scales is largely unknown, fine tuning is
generically required in order to obtain sufficient inflation
(while also allowing inflation to end at some point) and to
describe the small amplitude of the observed temperature
fluctuations in the CMB.

It is natural to look to string theory to explain the unnaturally
small numbers required by inflation, since it purports to describe
the relevant scales, and is arguably our only known consistent
description of gravity at the shortest distances. Unfortunately,
string realizations of inflation have proven hard to find, partly
due to the difficulty in identifying the inflaton from among the
plethora of string states, and partly due to the difficulty of
obtaining de Sitter solutions to the low-energy effective field
equations. Recently, however, a string-based inflationary scenario, based on
brane-antibrane attraction and subsequent collision and
annihilation, was proposed \cite{classic}, which circumvents some
of these difficulties (see also \cite{others} for related work).
 The proposal has the following successes:
\begin{itemize}
\item
The relevant brane-antibrane interaction arises at string
tree-level, and so is explicitly calculable once the low-energy
spectrum of the model is known.
\item
Inflation naturally ends once the brane-antibrane separation is of
order the string scale, since at this point one of the open-string
modes becomes tachyonic, opening up a new direction along which
fields may roll. This furnishes a string-theory realization of
the hybrid-inflation scenario.
\item
An explanation is suggested for the dimension of the branes which
survive to late times, by attributing these to be the end products
of a cascade of earlier brane-antibrane annihilations.
\end{itemize}
This mechanism provides a geometrical realization of inflation in
terms of the position of branes and antibranes. It also hints at
new approaches to early-Universe cosmology, in which the initial
features of late-time cosmology are understood in terms of the
fundamental objects of string theory.

Of course, many questions also remain open within this scenario.
Probably the most pointed issue concerns the assumption that the
geometric moduli and dilaton field are frozen by some unknown
string physics at higher energies (perhaps the string scale).
Although it is certainly a logical possibility that these fields
are fixed at the string scale without affecting the rest of the
low-energy cosmology, no such construction has actually been made.

Next, the brane-antibrane collisions considered in \cite{classic}
only provided inflation for special initial conditions: parallel
branes colliding head-on starting from antipodal points within
purely toroidal compactifications. Although some of these special
conditions were argued in \cite{classic} to arise plausibly from
earlier brane evolution, others seem difficult to arrange in this
way.

Finally, the known purely toroidal compactifications do not lead
to phenomenologically realistic models with chiral matter, and so
are not of the most interest from a purely practical perspective
of being known to contain vacua which resemble our observed
low-energy world.

In this paper we propose an extension of the brane-antibrane
inflationary scenario, which extends the analysis of
\cite{classic} to address the open questions just mentioned. We
introduce two new changes in order to do so. First, instead of
using the inter-brane separation as inflaton, we now use one of
the extra-dimensional moduli for this purpose. Second, we
generalize the scenario from toroidal to orbifold
compactifications, motivated by the observation that these contain
a class of realistic string models. In particular, explicit
D-brane models within string theory have been constructed which
contain the standard model (or the left-right symmetric model)
spectrum among the massless states on the observable brane,
including realistic features such as three chiral families, proton
stability and gauge coupling unification \cite{bbarmod,aiqu}.

We find that brane-antibrane interactions within these models have
cosmological interpretations along the lines of
ref.~\cite{classic}, but with many improved features. Among these
are:
\begin{itemize}
\item
We find an almost generic occurrence of slow-roll inflation, with
the newly-liberated modulus of the extra dimensions as the
inflaton;
\item
We find that the small numbers required by inflationary scenarios
may be naturally explained without introducing numbers into the
underlying brane theory which are smaller than $\eps \sim 0.01$.
This happens because the cosmologically relevant quantities arise
as large powers of the small parameters of the underlying brane
physics. Better yet, the small quantities $\eps$ turn out to be
precisely the {\it same} small quantities which have been used to
explain the gauge hierarchy problem \cite{intermediate}.
\item
Inflationary exit occurs because of the intervention of stringy
physics, and not due to the violation of the slow-roll conditions
for the inflaton. Indeed some of the inflaton potentials we find would not
be considered viable if taken on their own, outside of string
theory, since they would appear to predict perpetual inflation.
\end{itemize}

These improvements are suggested by the following interesting
feature of the realistic models: consistency conditions (the
requirement of tadpole cancellations) force the existence of a
number of branes which are trapped at some of the singular
fixed-points of the orbifold, with another (different) number of
antibranes trapped at other orbifold fixed points.

Since these branes and antibranes are naturally trapped at the
fixed points, the separation between them can only be changed by
adjusting the size of the extra dimensions. Consequently we must
now imagine that this modulus is not fixed, and instead is light
enough to play the role of the inflaton field. Notice that the
interbrane separation and breathing modes have interchanged their
roles compared with ref.~\cite{classic}---the separation now being
fixed and the breathing mode being the inflaton. Our finding that
slow-roll inflation becomes generic follows from the change in
kinetic energy which is implied by this interchange.

Another interesting consequence of the tadpole-cancellation
conditions in these models is the requirement of a mismatch
between the number of branes and antibranes, with branes and
antibranes localized at different orbifold points. If the extra
dimensions shrink due to brane-antibrane attraction, it follows
that the endpoint cannot involve complete annihilation, since
conservation laws must require the existence of some remnant
branes. Once a realistic final state is identified at low energies
after inflation, this property may help understand why any branes
at all survive into the late universe.

The remaining sections are organized as follows. The next section
describes the brane configurations of interest, and identifies
their properties which are relevant to inflation. Section 3 then
describes the effective four-dimensional theory, and makes contact
with standard inflationary phenomenology. Our conclusions are then
briefly summarized in section 4. A calculation, which is used {\it
en passant}, is given in an Appendix.

\section{The Underlying String Model}

For concreteness we restrict our attention to Type-I, II string
configurations in which the six compact dimensions make up an
orbifold, which we take to be a product of three 2-dimensional
tori whose radii we denote by $r_i, i=1,2,3$. We keep track of
these three radii, and the dilaton, $e^\phi$, as the moduli which
are of potential cosmological interest at low energies.

\subsection{Branes at Orbifold Points}
In orientifold and orbifold models there are more types of BPS and
non-BPS branes than in toroidal compactifications. In particular
there are the so-called fractional branes which are simply branes
trapped at one fixed point of the orbifold. Following
the realistic orbifold constructions, we populate our spacetime
with D9 and D5 branes, and their antibranes, and we take the
5-branes to be fixed at various orbifold points in the extra
dimensions. (For some purposes it is convenient to also
consider orientifold planes.)
$T$-duality relates these configurations to those with D7 and D3
branes.

General tools exist for the construction of such models
\cite{ads}, where the trapping of branes and antibranes at the
singular fixed points is required by cancellation of tadpoles at
these points. In terms of the effective 4D theory the trapping of
the branes is needed in order to have anomaly cancellation,
therefore it is clear that otherwise the corresponding string
model would be inconsistent.

This class of models has been studied thoroughly during the past
few years and they offer a very interesting avenue towards
obtaining a realistic model from string theory. In particular it
has been shown that models with three families and the standard
model gauge group or its left-right generalization can be obtained
in this way and supersymmetry is broken by the presence of both
branes and antibranes \cite{bbarmod,aiqu}, which is realistic if
the corresponding string scale is intermediate
\cite{intermediate}. Furthermore these models allow the
unification of gauge couplings at that scale
\cite{bbarmod,dumitru}, providing the first explicit string
compactifications with gauge-coupling unification and proton
stability.

For concreteness, a typical model of this sort we have in mind is
a ${\bf Z}_3$ orientifold model with 27 fixed points. The standard
model lies inside a stack of 3-branes at one of the fixed points,
which are themselves immersed inside a stack of 7-branes (which
covers 9 of the fixed points). Cancellation of RR tadpoles then
requires also two stacks of anti 7-branes, each covering 9 other
fixed points, and separated from the 7-branes that contain the
standard model. This provides a realization of gravity-mediated
supersymmetry breaking with a realistic low-energy spectrum.
This picture has a (probably less clear)  $T$-dual
version in terms of the D5 and D9 branes with Wilson lines
\cite{aiqu}.

\EPSFIGURE[ht]{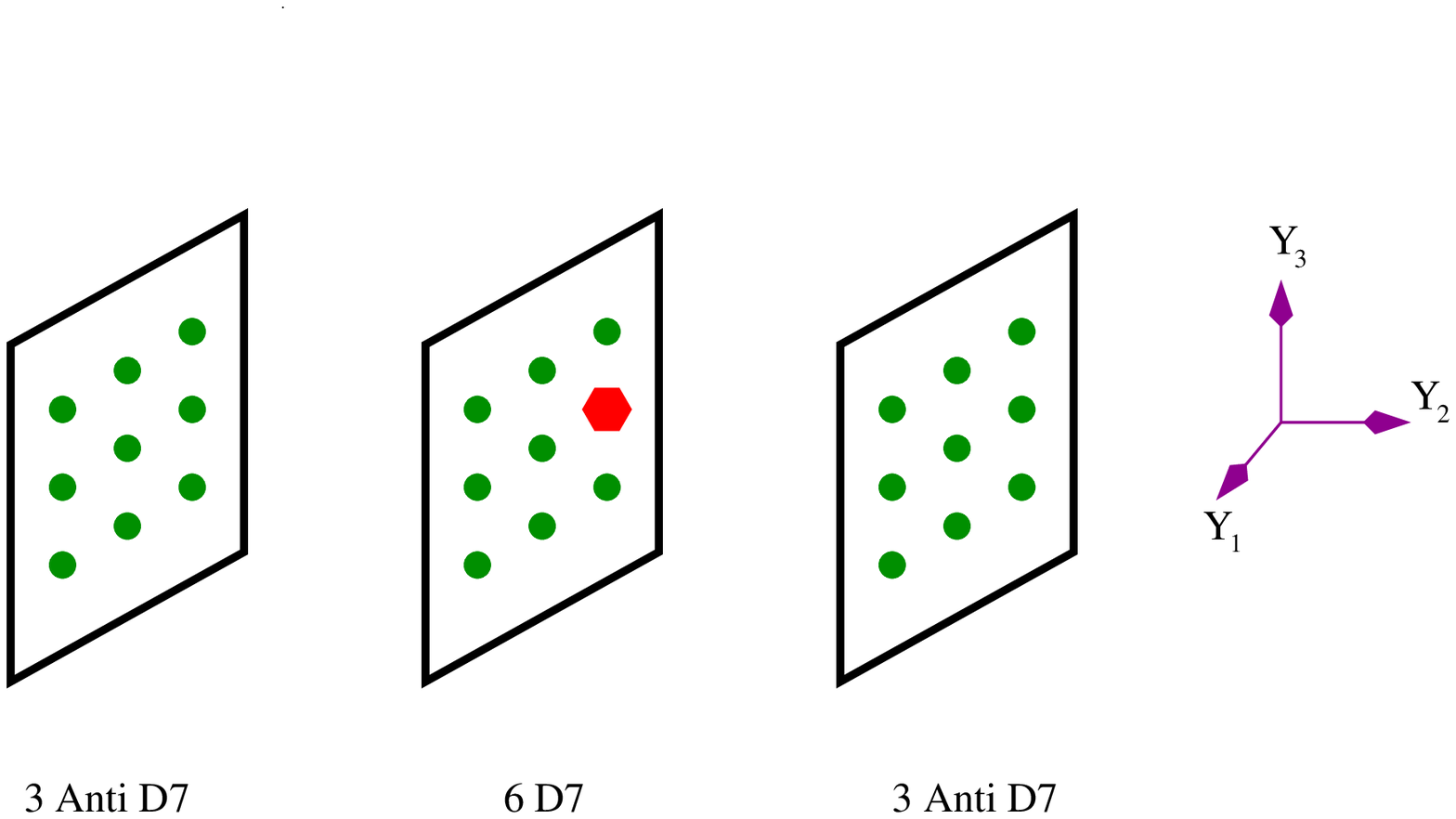,width=14cm} {A `realistic' model as in
reference \cite{aiqu}. Here we present the 6 extra dimensions in
terms of three complex coordinates $Y_i$. The standard model is at
a stack of D3 branes sitting in one of the 27 fixed points of the
${\bf Z}_3$ orbifold, represented here by the red hexagon. The
other fixed points are the green dots where other D3 branes may
live but the details are not important for us. The standard model
3 branes live inside a stack of 7-branes.
 There are also
anti 7-branes which are trapped at the fixed planes due to tadpole
cancellation conditions. The attraction of the branes and
antibranes generates a potential for the radius of the torus
transverse to the D7-branes similar to what it is considered in
the text.\label{figure1}}

Since the interbrane separation, $y$, is no longer a low-energy
degree of freedom, it is frozen within the effective theory. We
assume the same to be true of most of the other would-be moduli
and the dilaton (more about this later). Most importantly, we do
{\it not} imagine
 all of those moduli to be frozen, so at least one of them
appears within the low-energy effective theory, and the attractive
brane-antibrane potential should be regarded as being a function
of that modulus instead of $y$.

\subsection{Relevant Approximations}

Since we claim our results to be consistent with the low-energy
limit of string theory, there are several approximations which we
must make. We enunciate these approximations explicitly here,
because their failure signals the breakdown of the effective
field theory on which our inflationary analysis is based. In later
sections we look for situations where these approximations are
driven to fail, and interpret this failure as signalling the end
of inflation.

From the string point of view the two approximations we make are
the low-energy limit, and weak coupling. The first of these is
required because we shall work purely within the low-energy field
theory, and so require all of the relevant
distances---like the extra-dimensional sizes, $r_i$, and interbrane
separations---to be much larger than the string scale, $\ms r
\gg 1$. The weak-coupling approximation, $e^\phi \ll 1$, is
required to the extent that we use string perturbation theory to
identify the particle spectrum or to compute interbrane
interactions.

Within the low-energy (but higher-dimensional) effective field
theory, these approximations typically justify a classical
calculation using the linearized field equations. For instance,
recalling that $d$ denotes the number of compact dimensions, loop
effects in the field theory are controlled by the dimensionless
parameter
\begin{equation}
{G_d \over r^{d+2}} \sim {e^{2\phi} \over (M_s r)^{d+2}},
\end{equation}
where $G_d$ is the higher-dimensional Newton's constant. The
conditions $e^\phi \ll 1$ and $M_s r \gg 1$ ensure that this
parameter is small. Similarly, the approximation of working within
near-flat extra-dimensional geometries relies on the neglect of
the gravitational (and other) fields for which the various branes
act as a source. This neglect is controlled by the dimensionless
parameter
\begin{equation}
{G_d {\cal T}_p \over r^{d_\perp - 2}} \sim {e^{\phi} \, N\over (M_s
r)^{d_\perp -2}} ,
\end{equation}
where $d_\perp$ is the number of dimensions transverse to the
brane of interest, and we take the $p$-brane tension to be of
order ${\cal T}_p \sim N e^{-\phi}M_s^{p+1}$, 
with $N$ denoting the number of
coincident branes which source the fields. Unless $N$ is
excessively large, this quantity is also small if $e^\phi \ll 1$
and $M_s r \gg 1$.

Notice that $d_\perp -2 = d - p + 1 = (d+2)-(p+1) < d+2$ for all
$p>0$, so $M_s r\gg 1$ implies $G_d {\cal T}_p/r^{d_\perp -2} \gg
G_d/r^{d+2}$. It follows that as $r$ decreases it is the nonlinear 
corrections to the classical field equations which become important
before loop effects do. For a fixed $r$, the nonlinear classical effects
can always be made important without doing the same for loop
effects simply by making $N$ sufficiently large. Indeed, having multiple branes at
each point is not an uncommon occurrence within the realistic
orbifold models, but even for large $N$ 
our analysis using the linearized equations applies for sufficiently
large $r$.

\subsection{Inter-brane Interactions}
We now consider how the energetics of a multi-brane configuration
depends on the radii, $r_i$ and $e^\phi$, with an eye to finding
those terms which determine how these moduli evolve in the
low-energy theory. For widely-separated branes at weak coupling
this energetics is dominated by the brane tensions and by the
leading inter-brane interactions, which we now describe.

The exchange of bulk states which are approximately massless in
the higher-dimensional effective field theory dominates the
interactions of widely-separated branes at weak coupling. When the
interbrane separation itself is the inflaton \cite{classic}, these
interactions give the leading contribution to the inflaton
potential, since this is the leading term in the system energy
which depends on the inflaton. This is no longer the case if the
inflaton is a modulus, since then the interbrane interaction must
also compete with the brane tensions (as we discuss in more detail
below).

Consider for simplicity the attractive potential for a parallel
$p$-brane/antibrane pair separated by a distance $y$ which is
large compared to the string scale $1/\ms$. In an infinite
transverse space the interaction potential (per unit brane volume)
would be:
\eq \label{Potdef} W(y) =  -  {B \over y^{\dperp-2}}, \eeq
with
\eq \label{Bdefs} B = {\beta e^{2\phi}\over\ms^{2+d}}\,{\cal T}_p^2\,
\eeq
where $\dperp$ is the number of dimensions perpendicular to the
branes: $\dperp=d-p+3$ where $d$ is the number of compact extra
dimensions.

The constant $\beta$ is a dimensionless number which characterizes
the overall strength of the inter-brane force, and depends on the
number and types of bulk states which are effectively massless,
and so whose exchange generates a power-law potential. For
instance, direct D-brane calculation gives the result
corresponding to the exchange of the massless states of
10-dimensional supergravity ~\cite{Polchinski}, giving
\eq \label{kdef} \beta =   \pi^{-\dperp/2} \Gamma\left[
\frac{\dperp-2}{2} \right]\,. \eeq

The potential of eq.~\pref{Potdef} is not directly applicable to a
compact manifold like a torus (or orbifold). For instance, for a
square torus of size $r$ the periodic boundary conditions imply
the potential can be written as the infinite sum of image branes
at equivalent lattice points, giving \cite{classic}:
\begin{equation}
\label{naive} W({\bf y})\,=\,- \,\sum_{i}\,{B\over|{\bf y}-{\bf
y}_i|^{\dperp-2}}\,,
\end{equation}
where $\bf y$ gives the location of the antibrane and the sum is
over the location of the infinite image branes located at the
lattice points. A similar expression applies for more complicated
tori.

In what follows we wish to follow the potential as a function of
moduli like $r$, rather than the interbrane separation, $y$. In
the present instance this is most easily found by scaling $r$ out
of the coordinates, in which case eq.~\pref{naive} implies
\begin{equation}
\label{potencial} W(r)\ =\ - \frac{B G(\xi_*)}{r^{\dperp-2}}
\end{equation}
where $B$ are the same constants as before, and $G(\xi)$ is the
Green's function for a massless field on the torus expressed in
terms of the rescaled coordinates $\xi^i = y^i/r$. We imagine
$G(\xi)$ to be normalized so that it approaches $1/|\vec
\xi|^{d_\perp -2}$ in the limit that the brane and antibrane
approach one another: $\xi \to 0$.

Similar formulae also apply for other combinations of parallel
$p$-branes, such as a $D5$ brane and a parallel orientifold
5-plane, $O5$. The only difference in this case would be in the
precise value---and sign---taken by $B$. For our purposes this
value, and the value of $G(\xi_*)$, need not concern us, so long
as their product is ${\cal O}(1)$ and does not vanish. Ultimately
$B$ depends both on the type of branes involved and on the number
and types of massless particles whose exchange generates the
potential, and generally does not vanish except for
supersymmetric BPS-brane configurations. The value for $G(\xi_*)$
is also nonzero, and is computed explicitly in the Appendix for a
simple example of square torus, corresponding to a brane-antibrane
configuration on a $Z_2$ orbifold, as in Figure (2).

\EPSFIGURE[h]{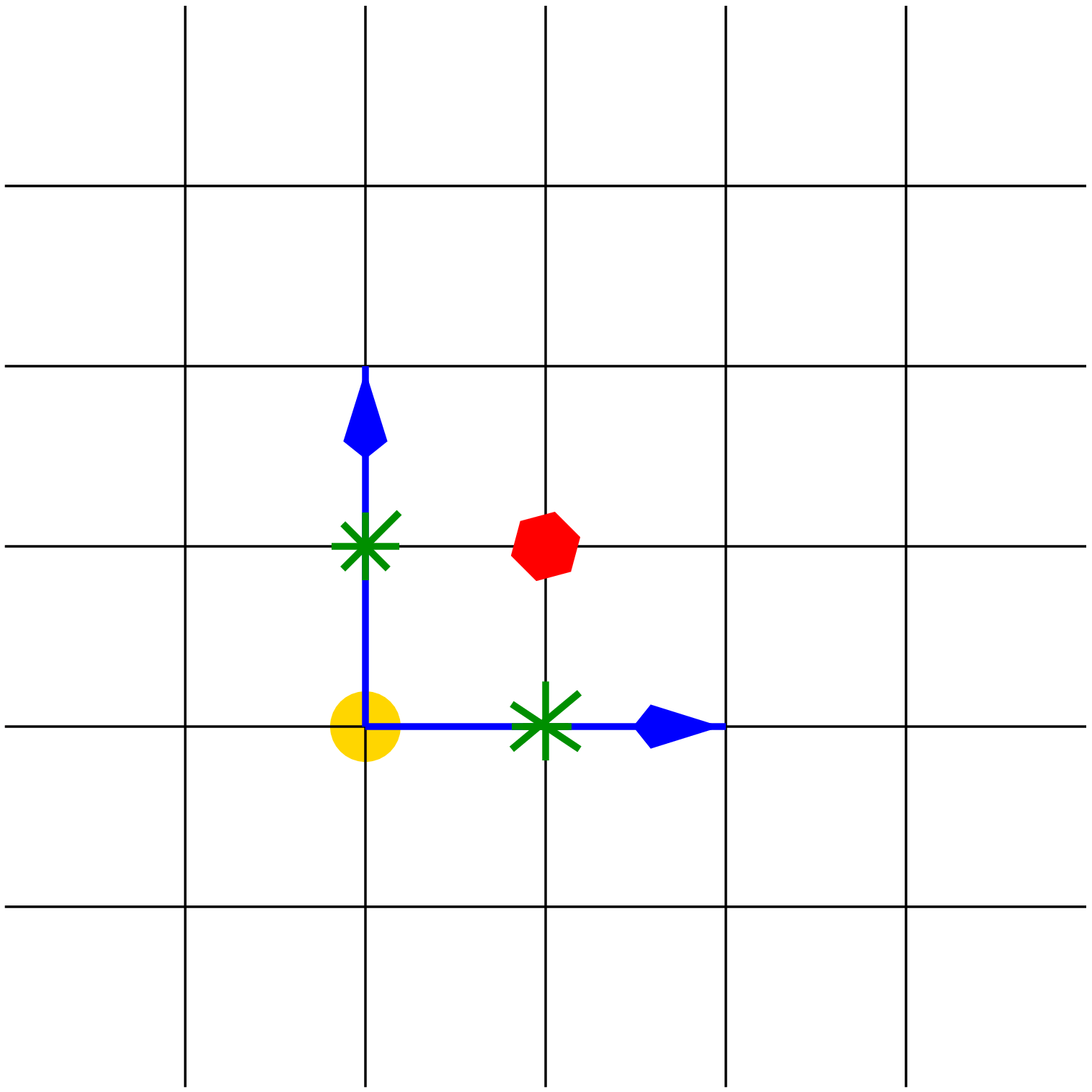,width=6.2cm} {A picture of the ${\bf
Z}_2$ orbifolds we are considering for cosmological purposes, with
branes and antibranes trapped at different fixed points. Of the 4
fixed points shown there the yellow circle represents a stack of
branes, the red hexagon a stack of antibranes and the other two
fixed points are empty.\label{figure2}}

Actually $G(\xi)$ also depends on the other moduli of the torus,
and leaving it as a free parameter allows us to consider how the
interaction energy depends on these variables. We emphasize that a
potential like (\ref{potencial}) should also apply to the cases of
most phenomenological interest, including the ${\bf Z}_3$ orbifold
of the six dimensional torus discussed earlier.

\section{The Effective 4D Theory}

We now turn to the effective 4D theory which describes the
cosmology of the model. The main assumption on which such a 4D
analysis rests is that inflation occurs on scales which are small
enough that Kaluza Klein (KK) modes are not excited as inflation
proceeds.

There is another, less obvious, condition for the validity of this
kind of 4D analysis which we must also impose. Besides working
within the 4D theory, we only allow a single field to be rolling
during inflation, and this is only justified to the extent that
the modulus we choose is the only 4D mode which is excited. In
particular, we must assume that other moduli are fixed by physics
at higher scales, such as at the string scale. If not, these other
moduli could also roll and we would have to re-evaluate whether
the universe's evolution is dominated by the scalar potential
rather than the kinetic energies of these other modes. Although
string cosmology is likely to involve more moduli than just the
single one which we liberate here, we regard our present
calculation to be the first (encouraging) step in relaxing the
frozen-moduli assumptions that are implicit in the extant string
cosmology proposals, including that ref.~\cite{classic}.

With these {\it caveats} in mind, we require the 4D effective
theory which describes the low-energy dynamics of the brane
configuration whose dynamics is of interest. Consider for this
purpose the 6-torus described earlier. The brane configuration of
interest has two parts. We imagine first having $N_9$ $D9$ and
$\overline{D9}$ branes, which are stabilized against mutually
annihilating by some mechanism. (This might be accomplished, for
instance, if there were Wilson lines within the $D9$ but not the
$\overline{D9}$, leading to these two 9-branes not carrying
precisely opposite values for all conserved charges.) We
supplement these with a collection of $N_5$ parallel $D5$ branes
which are localized at an orbifold point of the four dimensions
whose radii are $r_2$ and $r_3$, and which wrap the two compact
$r_1$ directions. Parallel to these we also take $N_5$
$\overline{D5}$ branes which are located at another orbifold point
which is displaced relative to the first in the transverse four
dimensions.

\bigskip

\subsection{Identifying the 4D Moduli}
Although the radii, $r_i$, and dilaton, $e^\phi$, are moduli in
perturbation theory, we imagine that all moduli but one
combination of these is fixed by (unknown) physics at scales above
the compactification scale. Since the inflation we find in the
low-energy theory depends on which combination of moduli is not
fixed by this physics---leaving it free to play the role of the
inflaton in the low-energy theory---we pause briefly here to
motivate the choice we will make.

Since the compactification scales to which we will be led are much
larger than the electroweak scale, we expect the physics
responsible for stabilizing the moduli to respect (at the very
least) $N=1$ 4D supersymmetry in the large dimensions.
Consequently, the would-be moduli would naturally be grouped with
other scalars into complex fields, to form $N=1$ chiral
supermultiplets. The way in which this happens is well understood
within weak-coupling string theory \cite{SusyMultsOld}, and in the
present case $r_i$ and $e^\phi$ combine into the real parts of
four (dimensionless) complex scalar fields, $S$ and $T_i,
i=1,2,3$, according to \cite{SusyMultsNew}
\eq s := {\rm Re}\; S = e^{-\phi} \, M_s^6 \; r_1^2 r_2^2 r_3^2,
\qquad  t_i := {\rm Re}\; T_i = e^{-\phi} \, M_s^2 \; r_i^2. \label{stdef}\eeq
Notice that the conditions of weak coupling ($e^\phi \ll 1$) and
large spatial dimensions ($M_s \, r_i \gg 1$) imply that the
moduli satisfy $s, t_i \gg 1$.

The simplest way to see that these combinations transform as the
real part of a chiral scalar is to recognize that this combination
of radii and dilaton is what appears as the coefficient of the
tree-level gauge kinetic terms for gauge bosons which propagate on
the 9-branes and 5-branes, or on 5-branes which wrap the other
four compact directions, if these had been present.

Now comes the main point. All we assume about the dynamics which
would fix these moduli is that it occurs well above the
effective  $N=1$
supersymmetry breaking scale, and so is likely to constrain the
entire complex scalars, $S$ and $T_i$. Furthermore, since these
fields typically represent the gauge couplings of different gauge
groups, each of which might get strong at different scales, it is
plausible to take the inflaton to be simply one of the four
fields, $s$ or $t_i$ (as opposed to choosing simply $e^\phi$ or
one of the $r_i$'s, for instance). We shall find in what follows
that if this is true, then the resulting 4D effective potentials
for the remaining modulus quite naturally give rise to inflation
with many attractive features.

\subsection{The 4D Modulus Potential}

We next compute the low-energy scalar potential, $V$, for the
fields $s$ and $t_i$ within the effective 4D theory. We shall do
so as a function of all four of these fields, in order to explore
which might be the best inflaton, keeping in mind that all of the
others are to be considered fixed by higher-dimensional dynamics
once the inflaton is chosen.

The contributions of the two main sources for the potential at
weak coupling are the various brane tensions, and their mutual
interactions. Once integrated over the extra dimensions, the
tension of a $D9$ or $\overline{D9}$ brane contributes the
following to the four-dimensional potential:
\eq \label{9Tension} V_9 = {\cal T}_9 \; r_1^2 r_2^2 r_3^2, \eeq
where the 9-brane tension is ${\cal T}_9 = \alpha_9 M_s^{10}
e^{-\phi}$. Here $\alpha_9$ is a dimensionless constant whose
precise value is not important for our purposes.

By contrast, the tension of a $D5$ or $\overline{D5}$ brane
depends only on the volume of the two compact dimensions which are
wrapped by the brane. For branes wrapping the two directions whose
volume is $r_i^2$ we have for the 4D potential:
\eq \label{5Tension} V_{5_i} = {\cal T}_5 \; r_i^2, \eeq
where the 5-brane tension is ${\cal T}_5 = \alpha_5 M_s^6
e^{-\phi}$.

These contributions may be summed to find the result for more
complicated brane configurations. For instance for a configuration
of $N_9$ $D9$-branes (and antibranes), with $N_{5_i}$ $D5$-branes
(and antibranes\footnote{
Notice we do not necessarily assume equal numbers of branes and
antibranes in these and later expressions.
} ) wrapping the $r_i$ directions, we would have:
\eq \label{ScPot} V_{\scriptscriptstyle T} = N_9 V_9 +
\sum_{i=1}^3 N_{5_i} V_{5_i} = {N_9 \, k_9 \; s} +  \sum_{i=1}^3
N_{5_i} \, k_{5_i} \; t_i ,  \eeq
where $k_{5_i}$ and $k_9$ are $O(1)$ dimensionless numbers. For
later purposes we remark that although $k_{5_i}$ and $k_9$ are
generically positive, they can be negative or zero for particular
brane configurations. For instance $k_{5_i}=0$ for equal numbers
of coincident $D5$-branes and orientifold planes ($O5$-branes)
wrapping the $r_i$ directions, because the positive tensions of
the $D5$-branes precisely cancel the negative tensions of the
$O5$-branes. In precisely the same way $k_9=0$ for the open
string, which corresponds to 32 $D9$-branes and a single $O9$
plane.

To eq.~\pref{ScPot} should be added the contribution of any
relevant inter-brane interactions, $V_{\rm int}$, as obtained by
summing the pairwise interactions (such as given by
eq.~\pref{potencial}) among the various branes. For example, the
4D dimensional reduction of the interaction between the $N_5$
coincident $D5$ branes and the $N_5$ coincident (but displaced)
$\overline{D5}$ branes would be
\eq \label{IBPV} V_{\rm int} = N_5^2 W(r) \; r_1^2 =\ - \left({
N_5^2 k_{\rm int} r_1^2 \over a r_2^2 + b r_3^2} \right) = -
\left( {N_5^2 k_{\rm int} t_1 \over a t_2 + b t_3} \right) \, ,
\eeq
where $k_{\rm int}, a$ and $b$ are dimensionless $O(1)$ numbers
which depend on the type of branes involved, and on their relative
position in the transverse dimensions.

\subsection{Kinetic Terms}
The potential just derived cannot be directly used to determine
how the modulus evolves within the low-energy theory because the
kinetic terms of the moduli and four-dimensional metric are not
yet in canonical form. In particular, non-minimal metric-modulus
couplings of the form ${\cal L} = F(s,t_i) \; \sqrt{-g} \, R$ can
compete with the scalar potential if the metric is not flat. An
analysis of the cosmology must be done {\it after} Weyl rescaling
the metric to put the Einstein-Hilbert action into standard
form.\footnote{We thank Jim Cline for reminding us of this fact.}

The kinetic terms for the metric and the moduli are obtained by
compactifying the ten-dimensional dilaton and Einstein-Hilbert
kinetic terms, which are\footnote{We use a positive-signature
metric, and follow Weinberg's curvature conventions.}
\eq \label{CMAction} {\cal L}_{10} = M_s^8 \, e^{-2\phi}
\sqrt{-{\cal G}} \; {\cal G}^{MN}  \left[ - \; \frac12 \, {\cal
R}_{MN} + 2 \,
\partial_M \phi \partial_N \phi + \cdots \right] \, .
\eeq

For compactifications our interest is in metrics of the form
\begin{equation}
\label{MetricForm} {\cal G}_{\scriptscriptstyle MN} =
\pmatrix{g_{\mu\nu}(x) & 0 & 0 & 0 \cr 0 & r_1^2(x) \, \delta_{mn} & 0
& 0 \cr 0 & 0 & r_2^2(x) \, \delta_{ab}& 0 \cr 0 & 0 & 0 & r_3^2(x) \,
\delta_{uv} \cr} ,
\end{equation}
where none of the internal metrics depend on the noncompact
coordinates $x^\mu$. The kinetic terms for the $r_i$ are found
using the result
\begin{equation}
{\cal G}^{MN}{\cal R}_{MN}=g^{\mu\nu}R_{\mu\nu}+4\sum_{i=1}^3 \nabla^2\ln r_i
+2\sum_{i=1}^3 \partial^\mu \ln r_i\partial_\mu r_i+4\sum_{i,j=1}^3
\partial^\mu\ln r_i\partial_\mu \ln r_j\,,
\end{equation}
where the covariant derivatives are with respect to the 4D
metric, $g_{\mu\nu}$.

By a Weyl rescaling $g_{\mu\nu} \to \lambda \; g_{\mu\nu}$
with $\lambda = M_p^2 e^{2\phi}/(M_s^8 \; r_1^2 r_2^2 r_3^2)$,
one can recast the reduced 4D Lagrangian in the standard no-scale form
\begin{equation}
{\cal L}_4=M_p^2\sqrt{-g}g^{\mu\nu}\left[-\frac{1}{2}R_{\mu\nu}
-\frac{1}{4}\partial_\mu\ln s\partial_\nu\ln s-\frac{1}{4}\sum_{i=1}^3
\partial_\mu\ln t_i\partial_\nu\ln t_i\right]+\cdots\,,
\end{equation}
where $s, t_i$ are defined in Eq. (\ref{stdef}).

The Einstein-frame scalar potential is obtained from our previous
expressions by performing this rescaling. For instance, the
potential to which this leads for $N_9$ $D9$-branes and antibranes
plus $N_{5_1}$ parallel $D5$-branes and $\overline{D5}$-branes
wrapping the $r_1$ dimensions would be
\eq \label{EFPot} V = \lambda^2 \Bigl[ V_{\scriptscriptstyle T} +
V_{\rm int} + \cdots \Bigr] = M_p^4 \left[ {k_9 N_9 \over t_1 t_2
t_3} + { N_{5_1} \over s t_2 t_3} \left( k_{5_1}
 - {k_{\rm int} N_{5_1} \over a t_2 + b t_3}\right)
 + \cdots \right]\; . \nonumber \eeq
Notice that---provided $N_{5_1}$ is $O(1)$---the interaction
terms are suppressed compared to the tension terms when $t_2$ and
$t_3$ are large, as would be expected for well-separated branes at
weak coupling.

\subsection{An Einstein-Jordan Dictionary}
Since it can be confusing to identify the important mass scales
when passing between Jordan (string) and Einstein frames, we here collect a
few relevant expressions. The main scales in the problem are:
($i$) the 4D Planck scale, $M_p$, as defined by measurements of
the strength of gravitational interactions in the low-energy 4D
theory; ($ii$) the string scale, $M_s$, as defined, say, by the
masses of closed-string states in the bulk; and ($iii$) the
Kaluza-Klein scale, $\mu_{\scriptscriptstyle KK_i}$, as defined by
the masses of KK excitations in the $r_i$ dimensions. The
expressions for these in both the Jordan and Einstein frames are
listed in Table 1, where $M$ is the mass scale that appears in the
Jordan, or string, frame and Einstein frame actions.

\begin{center}
\begin{tabular}{ccc}
Mass  &  \qquad Jordan Frame &  \qquad Einstein Frame \\
\hline\\
$M_p$ & \qquad $M^4 e^{-\phi} \, r_1 r_2 r_3$ &  $M$ \\
$M_s$ & \qquad $M$ & \quad $\left( M^2 e^{-\phi} r_1 r_2 r_3
\right)^{-1}$ \\
$\mu_{\scriptscriptstyle KK_i}$ & \qquad ${2 \pi \over r_i}$ & \qquad $\left( {2
\pi \over r_i} \right) \, \left( M^3 e^{-\phi} r_1r_2r_3
\right)^{-1}$
\end{tabular}
\end{center}
\begin{center}
 {\small Table 1: The relation between physically-measured
masses \\
and the parameter $M$ in the Jordan and Einstein frames.}
\end{center}

From this expression we see that the expressions for all mass
ratios are the same in both frames, as must be true for all
physical quantities. In particular they are related to the
dimensionless quantities $s$ and $t_i$ by
\eqa \label{Ratios} {M_s \over M_p} &=& {1 \over M^3 e^{-\phi}
r_1 r_2 r_3} = \left( {1 \over s \, t_1 \, t_2 \, t_3}
\right)^{1/4} \nonumber \\
{\mu_{\scriptscriptstyle KK_1} \over M_s} &=& {2 \pi \over M
r_1} =  2\pi\; \left( {t_2 \, t_3 \over s \, t_1} \right)^{1/4}, \qquad etc.
\eeqa

\section{4D Cosmology}
With the Einstein-frame scalar potential we are in a position to
investigate the low-energy 4D cosmological implications of the
underlying brane dynamics. We imagine therefore a configuration of
9- and 5-branes (and antibranes), with the 5-branes wrapping the
three 2-tori which make up the compact six dimensions.

As we have mentioned before we restrict ourselves to the case
where we assume that all moduli are fixed by some string-theory
effect, with the exception of one which would be the inflaton
field. Notice that if we consider the potential as a function of
all the fields it runs away to zero string coupling and infinite
radii (the decompactification limit). This is somewhat
counterintuitive given the fact that the branes and antibranes
might be expected to attract one other, and it would seem natural
to think that they would make the transverse dimensions contract
rather than expand. This is a subtlety related to the fact that
one must draw such conclusions working within the Einstein frame
rather than the Jordan frame. This situation changes very much if
some of the moduli are frozen as we will now see.

\subsection{The Generic Situation}
Suppose $x = s$ or $t_i$ denotes the one modulus which is not
fixed by the higher-energy physics, and so which plays the role of
the inflaton. Then, considering only the tension terms in the
Einstein-frame scalar potential, we see that the inflaton potential
quite generically takes the following form:
\begin{equation}
\label{potX} V(X)\ = \kappa_0 + {\kappa_1 \over x} + \cdots =
\kappa_0 + \kappa_1 e^{-\sqrt{2} X/M_p} + \cdots \, ,
\end{equation}
where $\kappa_0$ and $\kappa_1$ are constants. The ellipses
here denote interaction terms and any others which fall off
more quickly for large values of $x$. $X$ is the
canonically-normalized field, which is related to $x$ by $x =
\exp(\sqrt{2} X/M_p)$.

In this section our goal is to summarize what these constants must
satisfy in order for this potential to exhibit slow-roll
inflation.\footnote{For a similar inflationary potential (with
$\kappa_1<0$) found in supersymmetric field theories see
\cite{stewart}.}

\bigskip\noindent{\bf Slow-Roll Conditions:}
Using standard definitions \cite{Rev}
\eq \epsilon = {\mpl^2\over2}\left({V'\over V}\right)^2\,,
\qquad\qquad \eta = \mpl^2\,{V''\over V}\,, \eeq
where the primes denote derivatives with respect to $X$, we find
\eq \epsilon\simeq \frac{\kappa_1^2}{\kappa_0^2}\ e^{-2\sqrt{2}
X/M_p} = \frac{\kappa_1^2}{\kappa_0^2 x^2} \,, \qquad\qquad
\eta\simeq \frac{2{\kappa_1}}{\kappa_0}\ e^{-\sqrt{2} X/M_p} =
\frac{2{\kappa_1}}{\kappa_0 x} \,, \eeq
provided $\kappa_0 \gg \kappa_1 /x$.

The 4D stress energy is potential dominated as long as the
slow-roll conditions ($\epsilon\ll 1, |\eta|\ll 1$) are satisfied,
which is the case provided only that $X$ is sufficiently large:
$X\gg M_p$. In terms of the original moduli we see that the slow
roll is equivalent to the requirement $x \gg 1$, which we have
already seen follows as a consequence of the use of the
weak-coupling, low-energy field-theory limit of string theory. We
see in this way that slow rolling is {\it generic} to all values
of the moduli which are large enough to allow the weak-coupling,
field-theory limit.

Since the energy density during this slow-rolling era is dominated
by $\kappa_0$, we see that inflation requires $\kappa_0 > 0$.

\bigskip\noindent{\bf Inflationary Exit:}
From the perspective of the four-dimensional theory, inflation
ends as soon as the slow-roll conditions are violated. For
instance, if $\kappa_1 < 0$, then the scalar potential drives $X$
to smaller values, eventually leading to an exit from the
slow-roll regime.

In our string-based picture this end of the slow roll is a
sufficient, but not a necessary condition for inflationary exit.
This is because inflation could also end if the approximations
which permit the 4D field-theory analysis should fail. This could
happen, for instance, if any of the $r_i$ should fall below the
string scale, or if the string coupling were to become strong.

Interestingly, we shall find that this second way of ending
inflation allows eq.~\pref{potX} to provide a realization of
inflation, even when $\kappa_1 > 0$. This is quite remarkable,
given that potentials of this form would not normally be
considered to be good for inflationary purposes because the slow
roll appears never to end. After all, in this case the potential
leads $X$ to {\it increase}, which puts it even deeper into the
slow-roll regime.

This is another example where knowing the underlying string
physics provides significant new insight. In the string case,
having $X$ grow typically requires some of the $r_i$ and/or
$e^\phi$ to shrink, thereby automatically driving the system
beyond the validity of our approximations, through the
introduction of new string degrees of freedom, or of strong
coupling effects. Without the knowledge of how $X$ relates to the
more fundamental quantities, $r_i$ and $e^\phi$, we would be blind
to the possible breakdown of the effective description, with the
subsequent potential for an exit from inflation.

\bigskip\noindent{\bf Sufficient Inflation:}
The number of $e$-foldings during inflation must satisfy
\cite{Rev}
\eq \label{efolding} N_e = \int_{X_h}^{X_{end}} dX \frac{V}{M_p^2
V'}\, \simeq 60 - \log\left( {10^{16} \; \hbox{GeV} \over
V^{1/4}_{\rm inf}} \right), \eeq
where $X_h$ is the field value at horizon exit and $X=X_{end}$
denotes the field value when inflation ends. $V_{\rm inf}$ denotes
the (constant) value taken by the scalar potential during
inflation, which for simplicity we also assume to be the energy
density after reheating.

Substituting the potential of eq.~\pref{potX}, this gives:
\eq \label{suffinf} x_{h} - x_{end} = - \; {2 \kappa_1 N_e \over
\kappa_0} = - \; {2 \kappa_1 \over \kappa_0} \left[ 60 - \log
\left( {10^{16} \; \hbox{GeV} \over \kappa_0^{1/4}} \right)
\right] \, ,\eeq
which is to be regarded as determining the value for $x_h$ at
horizon exit.

\bigskip\noindent{\bf Density Fluctuations:}
Fine tuning typically enters an inflationary scenario once it
attempts to explain the amplitude of the density perturbations
when they re-enter the horizon, as they are observed in the
present-day cosmic microwave background (CMB). The amplitude of
these primordial density fluctuations may be written as follows:
\eq \delta_H \sim {1\over 5\,\sqrt{3}\,\pi}\, \left( {V^{3/2}
\over \mpl^3\,V'}\right) \simeq  {1\over 5\,\sqrt{6}\,\pi}\,
\left( {\kappa_0^{3/2} x_h \over \mpl^2  \kappa_1} \right) \,,
\eeq
where $x_h$ again denotes the value of the relevant
modulus, $x$, at horizon exit. This agrees with the size of the
observed temperature fluctuations of the CMB if $\delta_H = 1.9
\times 10^{-5}$~\cite{COBE}.

The spectral index of the fluctuations is:
\eq \label{spectral} n-1=-6\epsilon_h +2\eta_h \simeq 2\eta_h
\simeq \frac{4{\kappa_1}}{\kappa_0 x_h} \, \eeq
where the subscript $h$ again denotes evaluation at horizon exit.
According to eq.~\pref{spectral}, the sign of $n-1$ is determined
by the sign of $\kappa_1$. Notice that the equation also
simplifies if $x_h \gg x_{end}$, which is possible when $\kappa_1
<0$, since in
this case $n-1$ can be written purely in terms of $N_e$ as
follows: $n-1 = -2/N_e$.

We next consider two situations in turn, depending on whether the
role of the inflaton is played by the modulus $s$ or $t_1$.

\subsection{$s$ -- Inflation}
Imagine first that all three $t_i$ are fixed by higher-energy
string physics, leaving only $x=s$ as the inflaton, and consider a
configuration of $N_9$ 9-branes (and antibranes) with $N_5$
5-branes and antibranes wrapping only the two $r_1$ directions.
The constants $\kappa_0$ and $\kappa_1$ in this case are given by
\eq \label{kappavals} \kappa_0 = {k_9 N_9 \mpl^4\over t_1 t_2 t_3}
\qquad \hbox{and} \qquad \kappa_1 = {\mpl^4 \over t_2 t_3} \;
\left[ k_5 N_5 - { k_{\rm int} N_5^2 \over a t_2 + b t_3} \right]
. \eeq
Clearly, positive brane tension requires $\kappa_0 > 0$, but
$\kappa_1$ can have either sign depending on the relative size of
the two terms in the square brackets.

Some care is required to handle the case where $\kappa_1$ is
negative. If $\kappa_1<0$ is accomplished by taking $N_5$ large,
then the approximation of using only single graviton (and dilaton
{\it etc.}) exchange to compute the interaction can break down,
since a full calculation of the metric (including the backreaction due to the
branes) of the transverse space may
be required. An alternative way to achieve negative $\kappa_1$
without this problem might be to place a coincident $D5-O5$ pair at
one orbifold point, with a coincident $\overline{D5} -
\overline{O5}$ parallel to it, but at a different orbifold point.
In this case cancellation of $D5$ and $O5$ tensions makes $k_5
\simeq 0$, allowing the attractive interaction to dominate without
the necessity of having $N_5$ be large.

Since the prospects for inflation depend on the sign of
$\kappa_1$, we now consider both signs in turn.

\subsubsection{$\kappa_1 > 0$}
For this case $s$ is driven during inflation to larger values.
Having $s$ grow with the $t_i$ fixed corresponds to having all of
the $r_i$ grow with their ratios, $r_i/r_j$, fixed, as well as
having $e^\phi$ grow so that $e^{-\phi} r_i^2$ does not change.
Although the $r_i$ never get driven down to the string scale, we
nevertheless are forced beyond the validity of our approximations
by the onset of strong string coupling when $e^\phi = O(1)$.
Here is an
example where inflation ends despite having the inflaton be driven
deeper and deeper into the slow-roll regime.

The slow-roll parameter becomes for this case:
\eq \label{suffinfs} \eta \simeq {2 \kappa_1  \over \kappa_0 s} \;
,\eeq
where, keeping in mind eq.~\pref{kappavals}, we have
\eq {\kappa_1\over \kappa_0} = {t_1 \over k_9 N_9} \; \left[ k_5
N_5 - { k_{\rm int} N_5^2 \over a t_2 + b t_3} \right], \eeq
and so $\kappa_1 /\kappa_0 \propto t_1 \gg 1$ if all other
constants are $O(1)$. As advertised, a slow roll is easy to obtain
since $\eta \sim t_1/s = 1/(M_s^2 r_2^2 r_3^2)$, which is always
much smaller than one given that $M_s r_i \gg 1$. 

To see how much fine-tuning is required within this scenario,
imagine we begin the inflationary evolution with all radii equal,
$r_1 = r_2 = r_3 = r_0$, and with  $e^{-\phi_0} \sim M_s r_0 \gg 1$. In this
case the time-independent moduli are $t_{i} = e^{-\phi_0} (M_s
r_0)^2 \sim e^{-3\phi_0}$, and we have initially $s_0 =
e^{-\phi_0} (M_s r_0)^6 \sim e^{-7\phi_0} \gg t_{i0} \gg 1$.

Since $\kappa_1 >0$, $s$ increases and inflation continues until
$e^\phi$ grows to $e^{\phi_{end}} \sim 1$, at which point we have
$(r_{i\; end}/r_{0})^2 \sim e^{\phi_{end}}/e^{\phi_0} \sim
e^{-\phi_0}$, so $(M_s r_{i\; end})^2 \sim e^{-\phi_0} (M_s
r_{0})^2 \sim e^{-3\phi_0}$. This means that the final value for
$s$ is $s_{end} = e^{-\phi_{end}} \prod_{i=1}^3 (M_s r_{i \;
end})^2 \sim e^{-9\phi_0} \gg s_0 \gg 1$. With these choices the
string scale (in Planck units) at the end of inflation
becomes---{\it c.f.} eq.~\pref{Ratios}---$(M_s/M_p)_{end} = (s_{end} t_1
t_2 t_3)^{-1/4} \sim e^{9\phi_0/2}$. The inflationary energy scale
is similarly, $\kappa_0^{1/4}/M_p = [k_9 N_9/(t_1 t_2 t_3)]^{1/4}
\sim (k_9 N_9)^{1/4} e^{9\phi_0/4}$.

If we drop the interaction term, involving $k_{\rm int}$, compared
with the tension terms, then the total number of $e$-foldings
which the noncompact dimensions undergo during the evolution from
$s_0$ to $s_{end}$ is
\eq N_{\rm tot} \sim {s_{end} \kappa_0 \over \kappa_1} \sim {k_9
N_9 \over k_5 N_5} \; \left( {s_{end} \over t_1} \right) \sim {k_9
N_9 \over k_5 N_5} \;  e^{-6\phi_0}, \eeq
which is larger than 60, provided only that $k_5 N_5 < k_9 N_9
e^{-6\phi_0}/60$.

Notice that the compact dimensions need not expand by anywhere
near as much as the noncompact, inflating dimensions do. For
instance, for the choices $k_5 N_5 \sim k_9 N_9$ and $e^{-\phi_0}
\sim 100$ the compact dimensions expand only by a factor $r_{i \;
end}/r_0 \sim 10$ as the noncompact dimensions expand by $N_{\rm
tot} \sim 10^{12}$ $e$-foldings.

The amplitude of primordial density fluctuations is:
\eq \delta_H \simeq  {1\over 5\,\sqrt{6}\,\pi}\,{\kappa_0^{3/2}
s_h \over \mpl^2 \kappa_1}  \sim  {(k_9 N_9)^{3/2} \over k_5 N_5
5\,\sqrt{6}\,\pi} \left[ {s_h  \over t_1^{3/2} (t_2 t_3)^{1/2}}
\right] \, . \eeq
where $s_h$ denotes the value of $s$ at horizon exit, whose value
we now must determine.

There are two cases to consider. First, imagine that horizon exit
occurs very close to the end of inflation, $s_h \sim s_{end} \sim
e^{-9\phi_0}$, as would happen if $N_{\rm tot} \gg 60$. Together
with $t_i \sim e^{-3\phi_0}$, this gives $s_h/[t_1^{3/2} (t_2
t_3)^{1/2}] \sim e^{-3\phi_0/2}$. We see the condition $\delta_H =
1.9 \times 10^{-5}$ becomes
\eq \label{sflucpos} {(k_9 N_9)^{3/2} \over k_5 N_5} \;
e^{-3\phi_0/2} \sim 7 \times 10^{-4}\, . \eeq
Since $e^{-\phi_0} \gg 1$ we see immediately that this scenario
requires $N_5$ to be unacceptably large. For instance, if we take
$k_5=k_9=1$, $e^{-\phi_0} \sim 100$ and $N_9 = 2$ (one brane and
one antibrane), then eq.~\pref{sflucpos} requires $N_5 \simeq 4
\times 10^6$. Since with this choice we have $N_5\sim t_2, t_3$ we
see that the approximation of keeping only linearized fields is starting 
to break down.

Alternatively, suppose $N_{\rm tot} \sim 60$, and so $s_h \sim s_0
\sim e^{-7\phi_0}$. In this case we have $s_h/[t_1^{3/2} (t_2 t_3)^{1/2}]
\sim e^{\phi_0/2}$, so the condition $\delta_H = 1.9 \times
10^{-5}$ becomes
\eq \label{sflucpos2} {(k_9 N_9)^{3/2} \over k_5 N_5} \;
e^{\phi_0/2} \sim  7 \times 10^{-4}\, . \eeq
A representative choice which satisfies these conditions, while
(just) remaining within the slow-roll regime is $N_9 = 2$,
$e^{-\phi_0} \simeq 6$ and $N_5 \simeq 1600$. With these choices
we also have $r_{i\; end}/r_0 \sim \sqrt6$, $(M_s/M_p)_{end}
\sim 3 \times 10^{-4}$ and $\kappa_0^{1/4}/M_p \sim 0.02$. Since
these choices imply $N_5\gg t_2, t_3$ we again see that $N_5$ is
unacceptably large.

\subsubsection{$\kappa_1 < 0$}
In this case the scalar potential drives $s$ to {\it smaller}
values. Since the $t_i$ are regarded as being fixed during this
evolution, we see that we must have all of the $r_i$ shrinking,
with their ratios $r_i/r_j$ fixed. At the same time $e^\phi$ also
shrinks in such as way as to keep $e^{-\phi} r_i^2$ constant. This
means that the evolution is towards weaker string coupling with a
smaller extra-dimensional radius. Once $M_s r_i = O(1)$ we leave
the domain of validity of our 4D analysis, and string physics
becomes relevant.

The evolution to small $r_i$ provides a natural stringy
realization of hybrid inflation \cite{hybridinflation}, in the
same spirit as proposed in refs.~\cite{classic,DT}. Unlike the
situation in ref.~\cite{classic}, we do not (yet) have a known
string instability on which to base the subsequent
post-inflationary evolution, although we mention some
possibilities in the next section.

For more precise estimates, we use again the scenario where all
radii start with sizes which are of order $M_s r_0 \sim
e^{-\phi_0}$. It follows that $t_i \sim e^{-3\phi_0}$ and the
initial value for $s$ is $s_0 \sim e^{-7\phi_0}$. 
Since $\kappa_1 <0$, $s$ decreases from this point
and inflation continues until $M_s r_{i\; end} = O(1)$. At this
point we have $(r_{i\; end}/r_{0})^2 \sim e^{2\phi_0}$ and so
$e^{\phi_{end}} \sim (r_{i\; end}/r_{0})^2 \; e^{\phi_0} \sim
e^{3\phi_0}$. The final value for $s$ in this case is $s_{end} =
e^{-\phi_{end}} \prod_{i=1}^3 (M_s r_{i \; end})^2 \sim
e^{-3\phi_0} \ll s_0$. The post-inflationary string scale (in
Planck units) becomes in this case $M_s/M_p \sim e^{3\phi_0}$.

Again using eq.~\pref{suffinfs}, and dropping the 5-brane tension
relative to the interaction term (as might be appropriate for the
$D5-O5$ configuration described earlier) gives
\eq N_{\rm tot} \simeq \left|{s_{0} \kappa_0 \over \kappa_1} \right|
\simeq {k_9
N_9 \over k_{\rm int} N_5^2} \; \left( {s_{0} (a t_2 + b t_3)
\over t_1} \right) \sim {k_9 N_9 \over k_{\rm int} N_5^2}\;
e^{-7\phi_0}, \eeq
which is again naturally very large.

Neglecting the 5-brane tensions compared to their attractive
interactions, the amplitude of primordial density fluctuations are
\eq \delta_H  \sim  {(k_9 N_9)^{3/2} \over k_{\rm int} N_5^2
5\,\sqrt{6}\,\pi} \left[ {s_h (a t_2 + b t_3) \over t_1^{3/2} (t_2
t_3)^{1/2}} \right] \, , \eeq
where again we must determine the value for $s_h$. Assuming for
the moment $N_{\rm tot} \gg 60$, then $s_h \sim s_{end} \sim
e^{-3\phi_0}$, so we find the condition $\delta_H = 1.9 \times
10^{-5}$ implies
\eq {(k_9 N_9)^{3/2} \over k_{\rm int} N_5^2} \; e^{3\phi_0/2}
\sim 7 \times 10^{-4}\, . \eeq

An attractive choice which gives the correct size for $\delta_H$
is $k_9 \sim k_{\rm int} \sim 1$, $N_9 = 2$ (one brane/antibrane
pair), $N_5 = 4$ (one $D5-O5$ and one
$\overline{D5}-\overline{O5}$ pair) and $e^{-\phi_0} \sim 100$.
Since this gives $N_{\rm tot} \sim 5 \times 10^{13}$ we have an
{\it a posteriori} justification for assuming $N_{\rm tot} \gg 60$.
These choices also imply $r_0/r_{i\; end} \sim e^{-\phi_0} \sim
100$, the string scale after inflation is $M_{s \; end} \sim
e^{3\phi_0} M_p \sim  10^{12}$ GeV and $\kappa_0^{1/4} = M_p/(t_1
t_2 t_3)^{1/4} \sim e^{9\phi_0/4} M_p \sim 10^{14}$ GeV.

The spectral index for these fluctuations satisfies $n < 1$, with
\eq |n-1| \simeq 2 \eta_h \simeq {4 k_{\rm int} N_5^2 \over k_9
N_9} \; \left( {t_1 \over s_h (a t_2 + b t_3)} \right) \sim {4
k_{\rm int} N_5^2 \over k_9 N_9} \; e^{3\phi_0} \sim 3 \times
10^{-5} ,\eeq
which is well within the experimental limits \cite{cmbexp}.

It is intriguing that this successful choice uses the same numbers
as does a natural brane-based solution to the hierarchy
problem \cite{intermediate}.

\subsection{$t_1$ -- Inflation}
A second obvious alternative is to use one of the $t_i$'s---say
$t_1$---as the inflaton, taking advantage of its absence in the
5-brane tension. In this case, if we ignore the interbrane
interaction term, for a collection of 5-branes and antibranes
wrapping only the $r_1$ directions we have an Einstein-frame
potential of the form of eq.~\pref{potX}, but with
\eq \kappa_0 = {k_{5} N_{5} \mpl^4 \over s t_2 t_3} \qquad
\hbox{and} \qquad \kappa_1 = {k_9 N_9 \mpl^4 \over t_2 t_3} \, .
\eeq
This potential is of less interest for inflation, because the
condition $s/t_1 = M_s^4 r_2^2 r_3^2 \gg 1$ makes difficult the
possibility of obtaining a slow roll, which requires a small value
for
\eq \eta \simeq {2\kappa_1 \over \kappa_0 t_1} \simeq {2 k_9 N_9
\over k_5 N_5} \; \left( {s \over t_1} \right). \eeq
This can only be small enough if $k_5 N_5/k_9N_9 \gg s/t_1$, and
so again requires an enormous number of 5-branes compared to 9-branes.

Rather than pursuing this possibility, we instead imagine working
with the open string---which has 32 $D9$ branes and one $O9$
brane, and so has $k_9=0$---and with 5-branes which wrap more
than one 2-torus. In what follows we take $N_{5_1}$ 5-branes and
antibranes to wrap the $r_1$ directions and $N_{5_2}$ of them to
wrap the $r_2$ directions. In this case the tension terms in the
Einstein-frame scalar potential again have the form of
eq.~\pref{potX}, but with
\eq \kappa_0 = {k_{5_1} N_{5_1} \mpl^4 \over s t_2 t_3} \qquad
\hbox{and} \qquad \kappa_1 =  {k_{5_2} N_{5_2} \mpl^4 \over s t_3}
\, . \eeq
In this case the slow-roll parameter becomes
\eq \eta \simeq {2\kappa_1 \over \kappa_0 t_1} \simeq {2 k_{5_2}
N_{5_2} \over k_{5_1} N_{5_1}} \; \left( {t_2 \over t_1} \right).
\eeq
which can be small even if $N_{5_1} \sim N_{5_2}$ provided
$t_1/t_2 = (r_1/r_2)^2 \gg 1$.

For illustrative purposes, we consider the case $\kappa_1>0$, although
more detailed exploration of other possibilities is clearly of interest.
With this choice, we expect the evolution to be towards large values for
$t_1$, driving us deeper into the slow-roll regime. Since $t_2,
t_3$ and $s$ are held fixed by assumption, this requires $r_1$ to
grow while $e^\phi$, $r_2$ and $r_3$ shrink, with the quantities
$r_2^2/e^\phi$, $r_3^2/e^\phi$, and $e^\phi r_1^2$ all fixed.
Notice that these last conditions require that the products $r_1
r_2$ and $r_1 r_3$ also do not change. Inflation then ends once
$r_2$ and/or $r_3$ get driven down to the string scale.
This is a particularly intriguing scenario, since it allows a
hierarchy to develop between the sizes of various internal
dimensions while the noncompact dimensions inflate.

The number of $e$-foldings after horizon exit becomes
\eq N_e \simeq {\kappa_0 \over 2 \kappa_1} \; (t_{1\; end} -
t_{1\; h}) \simeq {k_{5_1} N_{5_1} \over 2 k_{5_2} N_{5_2}} \;
\left({t_{1\; end} - t_{1\; h} \over t_2} \right), \eeq
so the modulus at horizon exit, $t_{1 \; h}$, is determined by
\eq \label{suffinfst3} {t_{1\; end} - t_{1\; h} \over t_2} 
\simeq {2 k_{5_2} N_{5_2} \over k_{5_1} N_{5_1}} 
\left[ 60 - \log\left( {10^{16} \;\hbox{GeV} \over
\kappa_0^{1/4} } \right) \right] \, .\eeq

Consider again the scenario wherein inflation begins with all
radii equal, $r_1 = r_2 = r_3 = r_0$, and with a very modest
hierarchy of scales and couplings $e^{-\phi_0} \sim M_s r_0 \gg
1$. In this case the time-independent moduli are $t_{2} = t_{3} =
e^{-\phi_0} (M_s r_0)^2 \sim e^{-3\phi_0}$ and $s = e^{-\phi_0}
(M_s r_0)^6 \sim e^{-7\phi_0}$, and we have initially $t_{1\; 0} =
e^{-\phi_0} (M_s r_0)^2 \sim e^{-3\phi_0}$.

After inflation ends, $M_s r_{2\; end} = M_s r_{3\; end} = 1$, and
so $r_{2\; end}/r_0 \sim e^{\phi_0}$. It then follows that
$e^{\phi_{end}} = (r_{2\; end} /r_0)^2 \, e^{\phi_0} \sim
e^{3\phi_0}$ and $r_{1\; end}/r_0 = r_0 / r_{2\; end} \sim
e^{-\phi_0}$. Clearly, then, $M_s r_{1\; end} = e^{-2\phi_0}$ and
so $t_{1\; end} =  e^{-\phi_{end}} (M_s r_{1\; end})^2 \sim
e^{-7\phi_0}$.

In terms of these quantities, the string scale at the end of
inflation becomes $(M_s/M_p)_{end} = (s t_{1 \; end} t_2
t_3)^{-1/4} \sim e^{5\phi_0}$, which is the TeV scale if
$e^{-\phi_0} \sim 2000$ and is the intermediate scale if
$e^{-\phi_0} \sim 40$. Notice that since $(\mu_{\scriptscriptstyle
KK_1} /M_s)_{end} \sim 2 \pi \; e^{2 \phi_0}$, for $e^{-\phi_0}
\sim 2000$ we have the intriguing values $M_s \sim 1$ TeV and
$\mu_{\scriptscriptstyle KK_1} \sim 2$ MeV. Although this is not
as low as is obtained in the ADD scenario \cite{ADD} (they are not
inconsistent because $e^{\phi_{end}} \ll 1$) it points to
potentially rich low-energy phenomena.

The amplitude of primordial density fluctuations is:
\eq \delta_H \simeq  {1\over 5\,\sqrt{6}\,\pi}\,{\kappa_0^{3/2}
t_{1\; h} \over \mpl^2 \kappa_1}  \sim  {(k_{5_1} N_{5_1})^{3/2}
\over k_{5_2} N_{5_2} 5\,\sqrt{6}\,\pi} \left[ {t_{1 \;h}  \over
t_2^{3/2} (s t_3)^{1/2}} \right] \, , \eeq
so the condition $\delta_H = 1.9 \times 10^{-5}$ gives
\eq \label{omegaexp}
{(k_{5_1} N_{5_1})^{3/2} \over k_{5_2} N_{5_2}} \; e^{\omega
\phi_0} \sim 7 \times 10^{-4}\, . \eeq
Here $\omega = 13/2$ if we take horizon exit to occur for $t_{1\;
h} \sim t_{1\; 0} \sim e^{-3\phi_0}$. If, on the other hand,
horizon exit occurs towards the end of inflation, $t_{1\; h} \sim
t_{1\; end} \sim e^{-7\phi_0}$ then $\omega = 5/2$.

With the numbers we shall choose below $N_{\rm tot}
\sim N_e \sim 60$ and so we take $t_{1\; h} \sim t_{1\; 0} \ll t_{1\;
end}$. With this choice we have 
\eq N_e \sim 60 \sim {k_{5_1} N_{5_1}  \over 2 k_{5_2} N_{5_2} }
\; e^{-4\phi_0} \, ,\eeq
and $\omega = 13/2$ in eq.~\pref{omegaexp}.

A choice which satisfies these conditions would be $k_{5_1} \sim
k_{5_2} \sim 1$, $N_{5_1} = N_{5_2} = 2$ and $e^{-\phi_0} \sim 3$,
for which $r_{1\; end}/r_0 \sim e^{-\phi_0} \sim 3$,
$(M_s/M_p)_{end} \sim e^{5\phi_0} \sim 4 \times 10^{-3}$ and
$\kappa_0^{1/4}/M_p = (s t_2 t_3)^{1/4} \sim e^{13\phi_0/4} \sim
0.03$.

The spectral index obtained in this case is positive (since
$\kappa_1>0$) but is too large, for much the same reasons 
as for $s$-inflation with $\kappa_1 > 0$. We have
\eq n - 1 \simeq 2 \eta_h \simeq {4 k_{5_2} N_{5_2} \over k_{5_1}
N_{5_1}} \; \left( {t_2 \over t_{1\; 0}} \right) \sim  {4 k_{5_2}
N_{5_2} \over k_{5_1} N_{5_1}} \, , \eeq
which is not small because $t_{1\; 0} \sim t_2$ and $N_{5_1} \sim
N_{5_2}$. 

If instead we assume horizon exit occurs for $t_{1\; h} \sim t_{1 \; end}$, 
then the spectral index is still positive (since $\kappa_1>0$) but now has size
\eq n - 1 \simeq {4 k_{5_2} N_{5_2} \over k_{5_1}
N_{5_1}} \; \left( {t_2 \over t_{1\; end}} \right) \sim  {4 k_{5_2}
N_{5_2} \over k_{5_1} N_{5_1}} \, e^{4\phi_0}. \eeq

An attractive choice of parameters in this case takes
$k_{5_1} N_{5_1} = k_{5_2} N_{5_2} = 2$ and $e^{-\phi_0} \sim 20$,
which implies the string scale before and after inflation is
given by $(M_s)_0 \sim e^{4\phi_0} M_p \sim 10^{12}$ GeV and 
$(M_s)_{end} \sim e^{5\phi_0} M_p \sim 10^{11}$ GeV.
The spectral index is now well within the experimental limits. 


Once again, we obtain sufficient inflation and a correct amplitude
of primordial fluctuations without dialing in unreasonably
small parameters into the model.

\subsection{Tachyonic Phase Transitions and Inflationary Exit}

Brane realizations of inflation provide an extremely natural way
to end inflation, provided that inflation ends with the breakdown
of the low-energy field-theory approximation. In this case an
enormous number of string states become relevant, and can dramatically
change the cosmological evolution, providing a very elegant realization
of hybrid inflation \cite{hybridinflation}.

Ref.~\cite{classic} provided the first explicit example of this
mechanism using a known string instability when branes and antibranes
approach one another. The instability is due to the fact that at a
critical interbrane separation an open string state, with an endpoint on
each of the branes, becomes tachyonic. Physically, this instability
corresponds to the mutual annihilation of the brane and antibrane,
as has emerged from recent studies of tachyon condensation in string
theory following the original ideas of \cite{sen}.

In the present situation, there are also tachyonic states appearing at
critical distances, once $\ms r \sim 1$. However the physics of
the corresponding instability is very different, as might be
expected from the observation that
the numbers of initial branes and antibranes are typically not equal
in realistic models. At the critical radius the tachyon field which
develops corresponds to a direction in field space which does not lead
to the closed string vacuum, but to another non-BPS brane configuration.

The precise configuration to which the system evolves depends on the
details of the orbifold model and on how many dimensions are
being shrunk. For instance in representative examples
\cite{senmaj,abg,
Sen:1998ex}
(see also \cite{bogdan}), if the brane and antibrane approach one
another by having only one direction shrink, then at the critical
radius
 $r = r_c$
the charge and the tensions of the combined brane-antibrane pair
agree precisely with those of a single non BPS brane of one dimension
higher---usually referred to as a truncated brane---wrapped around
this direction. Since the tension of this truncated
brane decreases with the size of the
dimension that is shrinking this is a natural decay mode for the
brane-antibrane pair. So we can say that for $r>r_c$ energetics prefer
the brane-antibrane pair but for $r<r_c$ it is the single non-BPS brane
of one extra dimension which is stable.
For instance a pair of
D0-$\overline{\rm{D0}}$ branes on a contracting circle
decays to a single non-BPS D1 brane in type IIA
string theory.

 If two of the dimensions shrink then the $p$-brane/antibrane
decays into a $(p+2)$ brane-antibrane pair wrapping around the two
shrinking dimensions, and so on.
Notice that the fate of a 7-brane/antibrane pair after the shrinking
of the overall size of the extra dimensions decreases, is to decay into a
pair of 9-brane/antibranes, which do not annihilate with each other because
one of the branes carries a non trivial Wilson line.
The final result is that there are
different regions of parameter space where different brane-antibrane
pairs or non-BPS branes are stable. The detailed study of the regions
of stability and the full spectrum of BPS and non-BPS products depends
very much on the model.

\section{Discussion}

It is indeed striking that the models we consider can have such
promising cosmological features, including the much more generic
occurrence of inflation, with the extra bonus that they are based
on string configurations which are phenomenologically interesting.

We have found that slow-roll inflation becomes generic within the
low-energy field theory limit of string theory if the inflaton is
the radius of the compactified dimensions, whose shrinking is
driven by the tension and/or mutual attraction of various
branes and antibranes which are
localized at fixed points. The generic satisfaction of the slow-roll
conditions which
we find is in sharp contrast to what obtains for the same
interaction potential if the putative inflaton is the interbrane
separation in an internal space of fixed size.

We have also found configurations where the small inflationary
parameters which control the size of primordial density
fluctuations and the total amount of inflation are understood
without introducing by hand any numbers smaller than $0.01$. Even
better, the models to which we are led in this way are very
similar to the intermediate-scale string models
\cite{intermediate}, which use these same $O(0.01)$ numbers to
explain other small parameters (like the gauge hierarchy,
$M_w/M_p$, neutrino masses, {\it etc}).

The inflationary period very naturally ends as the size of the
contracting space reaches string sizes, at which point tachyonic
instabilities are known to arise, leading to the generation of new
brane configurations. This provides a new string-theory
realization of the hybrid inflation scenario. Furthermore, for
some moduli inflation ends in this way despite the fact that the
inflaton itself appears to remain deep within the slow-roll
regime. In this way string theory can resurrect low-energy
potentials which would otherwise have been discarded as not
providing an inflationary exit.

Some remarks are in order at this point concerning the consistency
between our inflationary cosmology and the well-known difficulties
obtaining de Sitter space as a solution to the supergravity field
equations \cite{nogo} (and so, by extension, to the low-energy
limit of string theory). The main point here is that the no-go
results are based on specific properties of the matter stress
energy, such as having a negative scalar potential, which are
satisfied by supergravity theories. These properties are simply
not satisfied by the potential energy which we find to drive
inflation.

Our potential need not satisfy these properties, despite arising
within a supersymmetric theory, because the branes themselves
break supersymmetry. Consequently, the low-energy theory of
fluctuations about the brane background only realizes
supersymmetry nonlinearly, having a low-energy particle content
which need not fill out linearly-realized supermultiplets for all
of the supersymmetries. This observation suggests a way to
interpret the no-go results. They may be thought to indicate that
the scale of inflationary physics should be below the
supersymmetry-breaking scale, within the effective theory within
which parts of particle supermultiplets have been integrated
out.\footnote{We thank Neil Turok for asking the question that led
to us to this argument, and Nemanja Kaloper for related
discussions.}

Even though the string-related phase transitions are not as well
studied as are the bulk brane-antibrane annihilations into the
closed-string vacuum, we can see what the differences might mean
for cosmology. First, since the minimum of the potential is not
the closed-string vacuum we may foresee that there is no risk that
all branes and antibranes completely annihilate, and so there may
no longer be a `branegenesis' problem, in which no branes one
which we might live survive into later epochs.

Next, to the extent that the final state will be a system of
branes, with a realistic spectrum of low-energy matter localized
on one of them, one might hope that the reheat energy would be
efficiently channelled into observable modes, and not frittered
away into unwanted bulk modes. If so, this would help with the
brane version of the cosmological moduli problem.

The precise amount of reheating which obtains depends on the
energy liberated by the transition, which is the difference
between the energy of the original brane pair and that of the
final state. It is clearly well worth determining this energy in
detail within a specific model.

We can say that the orbifold and orientifold models are richer than
the corresponding toroidal models for brane anti-brane cosmology.
Besides the attraction of stacks of fractional branes trapped at fixed
points that we have been discussing, there can be `bulk' branes
as in the toroidal case that can collide and reproduce the features
of \cite{classic}, (including the possibility of collisions between
stacks of $N$ branes colliding with stacks of $M$ antibranes with the
resulting $|N-M|$ branes surviving and absorbing part of the reheating
energy\footnote{We thank John March-Russell and Henry Tye for discussions on
this point.}). Alternatively,  there may be collisions of stacks of
 bulk branes with fractional branes that may happen in a way which
 does not ruin the tadpole cancellation \cite{bbarmod, aiqu},
 leaving again the remaining branes
trapped at singularities, with chiral matter---including the standard
model---absorbing part of the reheating temperature.

Although the scenario we propose here improves on many features of
the scenario of ref.~\cite{classic}, there is clearly still lots of
room to do better. We may include the effects of velocity dependence
on the potential as in \cite{tye}.
More importantly, even though we have relaxed the
assumption of freezing one of the moduli, we have not done the same for the
other toroidal moduli or the dilaton.

One way to improve our treatment would be to provide an explicit
construction of a mechanism for stabilizing these moduli. Of course
this is an old problem in string theory, and we have not yet been
successful in finding such a mechanism. As mentioned in the text we
may have the hope that nonperturbative effects can provide a
potential for the moduli that correspond to the gauge coupling on each
of the branes.
Recently there has been some progress in stabilizing precisely the
dilaton
field (although not so with the radius field)
 in terms of
Ramond-Ramond fluxes \cite{silverstein} or some string non-perturbative
effect (for recent discussions see for instance \cite{nonp}). In the
RR fluxes case we may need fluxes of more than one RR field to combine
with our potential and fix the dilaton without changing the rest of
our conclusions (the problem reduces to the one of finding a {\it dS}
solution from fluxes).
Addressing this issue in explicit models in a concrete way
  would be required in order to promote our
model into a definitive example of inflation within realistic string theory.

It would also be interesting to perform a similar analysis to the
other class of realistic type II models based on intersecting branes
\cite{intersectmod}.
 Even though these do not include anti-branes, they
are non supersymmetric and tend to have a tachyon in the spectrum
which can naturally realize the hybrid inflation scenario. 
For a recent effort in this direction see \cite{gbrz}. Similar
 discussions have  also appeared  in \cite{kallosh}.
Although much work remains to be done, we find the successes of the
present scenario very encouraging.

\acknowledgments

We acknowledge  J. Harvey, L. Ib\'a\~nez, 
 A. Lawrence, E. Martinec, N. Quiroz, M.
Rozali, B. Stefanski and especially A. Uranga for very useful
conversations. We thank J. Cline for pointing out an error in an
earlier version of this paper and H. Tye, J. Garc\'{\i}a-Bellido,
R. Rabad\'an,
J. March-Russell,  R. Blumenhagen, B. Koers, D. L\"ust and T. Ott
  for related communications. The
work of CPB was supported in part by N.S.E.R.C. (Canada) and
F.C.A.R. (Qu\'ebec); that of FQ by PPARC; GR was supported by DOE
grant DE-FG02-90ER-40560; and RJZ was supported by NSF grant
NSF-PHY-0070928.

\appendix

\section{Calculation of $G(\xi_*)$ on a square torus}

A calculation of the potential as a function of $r$ requires an evaluation
of the massless bulk-state propagator, $G(\xi)$, at the antipodal point of
the torus: $\xi^i = \xi_*^i = 1/2$ (using coordinates for
which $0\le \xi^i \le 1$). It turns out that expression \pref{naive}
is not the most convenient place to begin for this calculation. It is
not convenient because the sum in \pref{naive} diverges, with distant
images appearing to dominate the value of the potential.\footnote{ We thank 
Gary Gibbons for asking the question that motivated this discussion.}
Although this divergence does not complicate calculating the shape of
the potential as a function of $\xi$, it does preclude the simple calculation
of its value at $\xi=\xi_*$.

Since this divergence arises at large distances it has an infrared origin.
To see this consider the direct mode-function representation of the propagator:
\begin{equation}
\label{modesum}
G(\xi-\xi')\ =\ {\sum_n}' \frac{u_n(\xi) u^*_n(\xi')}{\lambda_n}
\end{equation}
where $u_n(\xi)$ are the eigenfunctions of $\nabla^2$ for the torus 
(sines and cosines) and $\lambda_n = - (2\pi)^2(n_1^2 + n_2^2 + 
\cdots + n_{d_\perp}^2)$ is the corresponding eigenvalue, where 
$n_k, k=1,\dots,d_\perp$ are integers. The prime on the sum indicates that
it does not include the zero mode $n_1 = n_2 = \cdots = n_{d_\perp} = 0$.

Unlike for the infinite-volume continuum, the exclusion of the zero mode is 
crucial in order to construct $G(\xi)$ on the torus. It is precisely the inclusion
of the zero-momentum modes in each term of the sum in eq.~\pref{naive} which
causes the problem with its convergence. 

To evaluate $G(\xi)$ it is simpler to directly
perform the mode sum in eq.~\pref{modesum}. This is most simply done
after first rewriting $1/\lambda_n = \int_0^\infty ds \; e^{\lambda_n s}$
(keeping in mind $\lambda_n < 0$),
after which the sums factorize into sums of the form:
\begin{equation}
\sum_n e^{- a n^2 + i b n} = \vartheta_3\left({b\over 2},e^{-a} \right),
\end{equation}
where $\vartheta_3(u,q)$ is the Jacobi $\vartheta$-function, and
we follow the conventions of ref.~\cite{WW}. In particular $\vartheta_3\left(
u,e^{-x}\right)$ has the asymptotic behaviors
\begin{eqnarray}
\label{asymptotic}
\vartheta_3\left( u,e^{-x} \right) &=& 1 + {\cal O}\left(e^{- x}\right), 
\qquad\qquad \hbox{as }\quad x \to \infty \nonumber\\
\vartheta_3\left( u,e^{- x} \right) &=& e^{u^2/x}\; \sqrt{{\pi \over x}}
\; \left[1 + {\cal O}\left(e^{-1/x}\right)\right], 
\qquad\qquad \hbox{as }\quad x \to 0.
\end{eqnarray}

Combining these expressions we find the following integral 
representation of $G(\xi)$:
\begin{equation}
G(\xi) = \int_0^\infty ds \left[ \prod_{k=1}^{d_\perp} \vartheta_3\left(
u_k, q\right) -1 \right] ,
\end{equation}
where $u_k = \pi \xi^k$ lies in the interval $(0,\pi)$ 
and $q = \exp[-4\pi^2 s]$. This expression
clearly converges for any $\xi\ne 0$, by virtue of the limiting forms, 
eqs.~\pref{asymptotic}.

Specializing to the antipodal point $\xi_*^k = 1/2$
gives $u_k = \pi/2$ for all $k=1,\dots,d_\perp$,
and so 
\begin{equation}
G(\xi_*) = \int_0^\infty ds \left\{\left[ \vartheta_3\left(
{\pi \over 2}, q\right) \right]^{d_\perp}-1 \right\}
= \int_0^\infty ds \left\{\left[ \vartheta_4\left(
0, q\right) \right]^{d_\perp}-1 \right\} .
\end{equation}
This expression is easily evaluated numerically, giving 
\begin{eqnarray}
G(\xi_*) &=& 0.01595,
 \quad\quad\hbox{for } \quad d_\perp = 3, \nonumber\\
G(\xi_*) &=& 0.01756 , \quad\quad\hbox{for } \quad d_\perp = 4, \\
G(\xi_*) &=& 0.01883, \quad\quad\hbox{for } \quad d_\perp = 5, \nonumber\\
G(\xi_*) &=& 0.01989 , \quad\quad\hbox{for } \quad d_\perp = 6, \nonumber.
\end{eqnarray}

\vskip 0.2in


\end{document}